\documentclass[doublespace]{article}

\usepackage{amssymb}
\usepackage{amsmath}   
\usepackage{amsfonts}
\usepackage{amsthm}

\newtheorem{thm}{Theorem}
\newtheorem{lem}[thm]{Lemma}
\newtheorem{prop}[thm]{Proposition}

\theoremstyle{definition}
\newtheorem{exmp}[thm]{Example}

\theoremstyle{remark}
\newtheorem{rem}[thm]{Remark}

\newcommand{\code}{\mathcal{C}}

\newcommand{\ff}{\mathbb{F}_q}

\newcommand{\fq}{\mathbb{F}_q}

\newcommand{\fqn}{\mathbb{F}_q^n}
\newcommand{\lgot}{\mathcal{L}}

\newcommand{\nat}{\mathbb{N}}

\newcommand{\rr}{\mathbb{R}}
\newcommand{\mr}{M_{\mathbb{R}}}

\newcommand{\zz}{\mathbb{Z}}

\newcommand{\codh}{\mathrm{H}^0 (X_P,\mathcal{O}(D_P))}

\newcommand{\talque}{~ | ~}
\newcommand{\ev}{\mathrm{ev}}

\newcommand{\citet}{\cite}

\newcommand{\citep}{\cite}

\newenvironment{pf}{\noindent{\it Proof}.}{\qed}

\title{On the Structure of Generalized Toric Codes}

\author{Diego Ruano\footnote{Partially supported by MEC MTM2004-00958 and FPU-AP2002-0087 and by Junta de CyL VA068/04, Spain. Address: Department of Algebra, Geometry and Topology, Faculty of Sciences, University of Valladolid, E-47005 Valladolid, Spain. E-mail: ruano@agt.uva.es}}

\begin{document}

\maketitle

\begin{abstract}
Toric codes are obtained by evaluating rational functions of a nonsingular toric variety at the algebraic torus. One can extend toric codes to
the so called generalized toric codes. This extension consists on evaluating elements of an arbitrary polynomial algebra at the algebraic torus
instead of a linear combination of monomials whose exponents are rational points of a convex polytope. We study their multicyclic and metric
structure, and we use them to express their dual and to estimate their minimum distance.
\end{abstract}

\section{Introduction}

J.P. Hansen introduced toric codes in \cite{ha}, these codes are algebraic-geometry codes at a toric variety over a finite field \cite{da}.
Algebraic-geometry codes are obtained by evaluating rational functions on a normal variety \cite{tsvl}. For a toric variety and a Cartier
divisor $D$, toric codes are obtained by evaluating rational functions of $\lgot (D)$ at the points of the algebraic torus $T=(\fq^\ast)^r$,
where $\fq$ is the finite field with $q$ elements. Toric codes have been studied in \cite{diguva,ha,ha3,jo,li2,li,ru}. In \cite{jo} there are
some examples of toric codes with very good parameters.

We extend the definition of toric codes to the so called generalized toric codes. Generalized toric codes are obtained by evaluating polynomials
at $T$ as for toric codes but considering arbitrary polynomial algebras instead of $\lgot(D)$. We emphasize that toric codes are generalized
toric codes. \citet{diguva} claimed that toric codes are multicyclic and it was proved there for a toric code defined using a toric surface. We
prove that generalized toric codes are multicyclic, and therefore toric codes coming from a convex polytope of arbitrary dimension. The aim of
this paper is to study the multicyclic and metric structure of generalized toric codes. We compute the dual of a generalized toric code, which
is a generalized toric code (the dual of a toric code is not a toric code in general). One cannot estimate its minimum distance using
intersection theory  \cite{ha3,ru} but we provide here a method to estimate the minimum distance similar to the one in \cite{li}  studying its
structure.

In the next section we have compiled some basics facts about toric codes and we also introduce the generalized toric codes. In section 3 we
study the multicyclic structure of generalized toric codes. Finally in section 4 we study their metric structure which makes it possible to
compute the dual of a generalized toric code. Furthermore we show that there are no self-dual generalized toric codes.

\section{Toric Codes and Generalized Toric Codes}

Let $M$ be a lattice isomorphic to $\zz^r$ for some $r \in \zz$ and $\mr = M \otimes \rr$. A convex polytope is the same datum as a toric
variety and Cartier divisor. Let $P$ be an $r$-dimensional convex polytope in $\mr$ and let us consider $X_P$ and $D_P$ the toric variety and
the Cartier divisor defined by $P$. We may assume that $X_P$ is non singular, in other case we refine the fan. Let $\lgot (D_P) = \codh$ be the
$\fq$-vector space of rational functions $f$ over $X_P$ such that $\mathrm{div}(f) + D_P \succeq 0$.

\vspace{0.2cm}

The \textbf{toric code $\code_P^t$ associated to $P$} is the image of the $\fq$-linear evaluation map
$$
\begin{array}{ccc}
  \ev:\lgot (D_P) & \to & \fqn \\
  f & \mapsto & (f(t))_{t \in T} \\
\end{array}
$$ where $T = (\fq^\ast)^r$. Since we evaluate at $\# T$ points, $\code_P^t$ has length $n=(q-1)^r$. For a toric variety $X_P$ one has that $\lgot (D_P)$ is the $\fq$-vector space generated by the monomials with exponents in $P \cap M$$$\lgot
(D_P) = \langle \{ Y^u = Y_1^{u_1} \cdots  Y_r^{u_r} \talque u \in P \cap M  \} \rangle \subset \fq [Y_1 , \ldots, Y_r ]$$

The dimension of the code and the kernel of $\ev$ are computed in \cite{ru}. Let $u \in P \cap M$ and $u = c_u + b_u$ where $c_u \in H = \{0,
\ldots, q-2\} \times \cdots \times \{0, \ldots, q-2\}$ and $b_u \in ((q-1)\zz)^r$. We will also denote $\overline{u}=c_u$. Let $\overline{P}= \{
\overline{u} \talque u \in P \cap M \}$. The dimension of the code $\code_P^t$ is $k= \# \overline{P}$.

The minimum distance of a toric code $\code_P^t$ is estimated using intersection theory \cite{ha,ru}. Also, it can be estimated using a
multivariate generalization of Vandermonde determinants on the generator matrix \cite{li}.

Let $U \subset H= \{ 0, \ldots, q-2 \} \times \ldots \times \{ 0, \ldots, q-2\}$, $T = (\fq^\ast)^r$ and $\fq [U]$ the $\fq$-vector space$$\fq
[U]=\langle Y^u = Y_1^{u_1} \cdots Y_r^{u_r} \talque u=(u_1, \cdots , u_r) \in U \rangle \subset \fq [Y_1, \cdots, Y_r]$$

\vspace{0.2cm}

The \textbf{Generalized toric code} $\code_U$ is the image of the $\fq$-linear map$$
\begin{array}{ccl}
  \ev: \fq[U] & \to     & \fq^{n} \\
  f           & \mapsto & (f(t))_{t \in T} \\
\end{array}
$$where $n = \# T = (q-1)^r$. Some of the results for toric codes are also valid for generalized toric codes. Namely, the following result ensures
that the map $\ev$ is injective and therefore the dimension of $\code_U$ is $k = \# U$.

\begin{lem} \label{le:ceCTG}
Let $U \subset H$ and set
$$
f=\sum_{u \in U} \lambda_u Y^u, \; \; \; \; \lambda_u \in \fq
$$
Then $(f(t))_{t\in T} = (0)_{t\in T}$ if and only if $\lambda_u = 0,\ \forall \  u \in H$.
\end{lem}

\vspace{0.4cm}

The proof of the previous result if the same as the one of \cite[lemma 3.2]{ru} for toric codes, and consequently we do not reproduce it. This
is because the proof for toric codes shows that a nonzero polynomial which is a linear combination of monomials of $H$ does not vanish
completely on $T$.

We have defined the generalized toric codes for $U \subset H$ as the evaluation of $\fq [U]$ at $T$.  As we claimed in the previous section,
this family of codes include the ones obtained evaluating polynomials of an arbitrary subalgebra of $\fq [Y_1 , \ldots, Y_r ]$ at $T$. The
following result shows this fact.

\begin{prop}\label{pr:nuCTG}

Let $V \subset \zz^r$, $\fq[V] = \langle Y^v  \talque v \in V \rangle$ and $\code_V$ the linear code defined by the image of the evaluation map
$\ev$ at $T$
$$
\begin{array}{ccl}
  \ev: \fq[V] & \to     & \fq^{n} \\
  f           & \mapsto & (f(t))_{t \in T} \\
\end{array}
$$

Let $v \in \zz^r$, where we write $v = c_v + b_v$ with $c_v \in H$ and $b_v \in ((q-1)\zz)^r$. We also denote it by $\overline{v}=c_v$. Then
$\code_U = \code_V$, where $U= \overline{V}$.

\end{prop}

\pf

Let $f = \sum_{v \in V} \lambda_v Y^v \in \fq[V]$ and $t \in T$. One has that$$f(t) = \sum_{v \in V} \lambda_v t^{c_v + u_v} = \sum_{v \in V}
\lambda_v t^{c_v}$$ And the result holds. \qed

Let $P$ be a convex polytope in $\mr$, by the previous proposition it follows that $\code^t_{P}= \code_{U}$ with $U=\overline{P}$. Therefore all
the results for generalized toric codes are valid in particular for toric codes.

\section{Multicyclic Structure of Generalized Toric Codes}

Multicyclic codes are those whose words are invariant under certain cyclic permutations, they can also be  understood as ideals in a certain
polynomial algebra. \citet{diguva} proves that a toric code defined using a plane convex polytope ($r=2$) is multicyclic by representing the
words of the code by matrices. The proof is hard to extend for arbitrary dimension because one should consider $r$-dimensional arrays, although
the result was claimed there for any $r$. We represent the words of the code by polynomials in order to prove that a generalized toric code of
arbitrary dimension is multicyclic.

Let $\code \subset \fqn$ be a linear code. We call $\code$ a \textbf{cyclic code} if $c = (c_0, c_1, \ldots , c_{n-1}) \in \code$ then $(c_{n-1}
, c_0 , c_1, \ldots, c_{n-2}) \in \code$.

Let $\fq[X]_{\le n-1}$ be the $\fq$-vector space of polynomials of degree lower than $n$ and $A$ the quotient ring $\fq[X]/(X^n -1)$. Since
$\fqn$, $\fq[X]_{\le n-1}$ and $A$ are vector spaces over the same field with the same finite dimension $n$ they are isomorphic. Then we
consider the isomorphisms
$$
\fqn \simeq  \fq[X]_{\le n-1}  \simeq \fq[X]/(X^n -1)
$$and for abbreviation one identifies $(c_0 , c_1 , \ldots , c_{n-1})$,     the polynomial $c_0 + c_1 X + \cdots + c_{n-1} X^{n-1}$ and the
class $c_0 + c_1 X + \cdots + c_{n-1} X^{n-1} + (X^n -1)$. In practice one uses the most convenient notation when no confusion can arise. A code
in the polynomial algebra $A$ is cyclic if and only if it is an ideal in $A$.

Cyclic codes have been deeply studied and used for real applications \cite{masl}. A natural extension of  cyclic codes are the so called
multicyclic codes. A code $\code \subset A= \fq [X_1 , \ldots , X_r] / (X_1^{N_1} -1 , \ldots X_r^{N_r} -1)$ is \textbf{multicyclic or $r$-D
cyclic} if it is an ideal in $A$, with $N_1, \ldots, N_r \in \nat$. Let $\fq [X_1 , \ldots , X_r]_{\le (N_1 -1 , \ldots, N_r -1)}$ be the
$\fq$-vector space of polynomials in the variables $X_1, \ldots , X_r$ of degree lower than $N_i$ in each variable $X_i$ for all $i$. In
particular, a cyclic code is a $1$-cyclic code. In the same way as for the cyclic case one can consider the following isomorphisms of vector
spaces
$$\fqn \simeq \fq [X_1 , \ldots , X_r]_{\le (N_1 -1 , \ldots, N_r -1)} \simeq A$$where $n = N_1 \cdots N_r$ and we can identify its elements.

Let $\code_U$ be the generalized toric with $U \subset H$. Set $\alpha$ a primite element of $\fq$, i.e. $\fq^\ast = \{\alpha^0, \alpha^1,
\ldots, \alpha^{q-2} \}$ and therefore $T = \{ \alpha^i = (\alpha^{i_1},  \ldots , \alpha^{i_r}) \talque i \in H \}$. Then $\code_U$ is the
vector subspace of $\fqn$ generated by $\{ (Y^u (\alpha^{i}))_{i \in H} \talque u \in U \}$, where $Y^u (\alpha^i) = \alpha^{\langle u,i
\rangle} = \alpha^{u_1 i_1+ \cdots + u_n i_n}$. In order to study the multicyclic structure we shall use the previous isomorphism, we denote the
code $\code_U$ in $A$ as $C_U^A$. Namely, we represent
$$
(\alpha^{\langle u,i \rangle})_{i \in H} \in \code_U ~ ~ ~ \mathrm{by} ~  ~ \sum_ {i \in H} \alpha^{\langle u,i \rangle} X^i \in \code_U^A$$

Let $U \subset H$ and $A = \fq[X_1, \ldots, X_r]/(X_1^{q-1} -1, \ldots, X_r^{q-1}-1)$. The code $\code_U^A \subset A$ which is isomorphic to
$\code_U \subset \fqn$ is
$$ \code_U^A = \{ \sum_{u \in U} \lambda_u \sum_ {i \in H} \alpha^{\langle u,i \rangle} X^i \talque \lambda_u \in \fq   \} \subset A$$

\begin{prop} \label{pr:ciclico}
Let $U \subset H= (\{0,\ldots,q-2 \})^r$, $\code_U^A$ is a $r$-D cyclic code with $N_1=q-1$, $\ldots$, $N_r=q-1$.
\end{prop}

\pf

Let $u \in U$, $\sum_{i \in H} \alpha^{\langle u , i \rangle} X^i \in \code^A_U$.

$ X^{a}\sum_{i \in H} \alpha^{\langle u,i \rangle} X^i = \sum_{i \in H} \alpha^{ u_1(i_1 -a_1) + \cdots + u_r (i_r -a_r) } X^i= \alpha^{-
\langle u,a \rangle}  \sum_{i \in H} \alpha^{\langle u,i \rangle} X^i $. And the results holds due to the linearity of $\code_U^A$. \qed

Besides of the product of polynomials in $\ff [H]$ which we denote by $\cdot$, $Y^u \cdot Y^v = Y^{\overline{u+v}}$, we consider the
multiplicative structure of $A$ in $\ff [H]$. The product of $A$ is given in the basis $\{ X^i \}_{i \in H}$ by $X^i * X^j = X^{i+j}$. The
following result pulls back the structure of $A$ in $\ff [H]$ which will be used in theorem \ref{th:mc}.

\begin{prop} \label{pr:prod}

Let us denote $\ev^{-1} (X^i)$ by $X^i$ in $\ff[H]$, then


\vspace{0.1cm}

$X^i * Y^u = \alpha^{- \langle u, i \rangle} Y^u$

\vspace{0.2cm}

$ Y^u * Y^v = \left\{ \begin{array}{ll} 0 & ~ \mathrm{if} ~ u \neq v\\
(-1)^r Y^u & ~ \mathrm{if} ~ u = v
\end{array} \right.$

\end{prop}

\pf

By the following isomorphisms considered above \begin{equation}\label{isomorfismo}
\begin{array}{rcccc}
\ff [H] &  \longleftrightarrow  & \fqn & \longleftrightarrow  & A \\
Y^u & \mapsto &(\alpha^{\langle u, i \rangle})_{i \in H} & \mapsto & \sum_{i\in H} \alpha^{\langle u, i \rangle} X^i
\end{array}
\end{equation}one has that

$X^i * Y^u = X^i * \sum_{j\in H} \alpha^{\langle u,j \rangle} X^j = \alpha^{- \langle u, i \rangle} Y^u$, by proposition \ref{pr:ciclico}.

$Y^u * Y^v =  \sum_{i\in H} \alpha^{\langle u,i \rangle} X^i * Y^v = \sum_{i\in H} \alpha^{\langle u - v ,i \rangle} Y^v=$ $$=\left\{ \begin{array}{ll} \sum_{i \in H} \alpha^{\langle u-v,i \rangle} Y^v = \frac{q (q-1)}{2}(\sup(u-v)   ) = 0 & ~ \mathrm{if} ~ u \neq v\\
\nonumber & \nonumber \\ \sum_{i \in H} Y^u = (-1)^r Y^u & ~ \mathrm{if} ~ u = v
\end{array} \right.$$where $\sup(u-v)$ is the number of nonzero coordinates of $u - v$.
\qed

The following result proves that any linear code over $\fq$ which is $r$-D cyclic with $N_1= q-1$, $\ldots$ , $N_r = q-1$, is a generalized
toric code. That is, the ideals of $A=\fq[X_1, \ldots, X_r]/(X_1^{q-1} -1, \ldots, X_r^{q-1}-1)$ are generalized toric codes. Therefore the
generalized toric codes and the $r$-D cyclic codes with $N_1= q-1$, $\ldots$, $N_r = q-1$ are the same family of codes.

\begin{thm} \label{th:mc}
Let $J \subset \fq[X_1, \ldots, X_r]/(X_1^{q-1} -1, \ldots, X_r^{q-1}-1)$ an ideal, then exists $U \subset H$ such that $J= \code_U^A$.
\end{thm}

\pf

Since $A$ is isomorphic to $\ff[H]$ by (\ref{isomorfismo}) and $\{ Y^u \talque u \in H\}$ is a basis of $\ff[H]$, we have that $\{ \ev(Y^u)
\talque u \in H\}$ is a basis of $A$, where $\ev(Y^u)=\sum_{i \in H} \alpha^{\langle u,i \rangle} X^i \in A$.

Let $\sum_{v \in H} \lambda_v \ev(Y^v) \in J$ and set $u \in H$, according to proposition \ref{pr:prod} we have that $\ev(Y^u)  \sum_{v \in H}
\lambda_v \ev(Y^v) = (-1)^r \lambda_u ev(Y^u) \in J$. Therefore $\ev(Y^u) \in J$ if $\lambda_u \neq 0$. We now apply this argument again, for
every generator of $J$ and $u$ in $H$, to obtain $U$ such that $J = ( \ev(Y^u) \talque u \in U )$. \qed

\section{Metric Structure of Generalized Toric Codes}

In this section we study the metric structure given by the bilinear form which defines the dual of a linear code, $\langle x , y \rangle =
\sum_{i=1}^n x_i y_i$ with $x,y \in \fqn$. The following result considers the metric structure of a generalized toric code $\code_U \subset
\fqn$ in $\fq[H]$ and computes its dual.

\begin{thm}\label{te:me}

With the above notations set $u, v \in H$, one has that

\vspace{0.1cm}

$ \langle \ev(Y^u) , \ev(Y^v) \rangle = \left\{
\begin{array}{ll} 0 & ~ \mathrm{if} ~ \overline{u + v} \neq 0\\
(-1)^r  & ~ \mathrm{if} ~ \overline{u + v} = 0
\end{array} \right.$

\vspace{0.1cm}

Let $u \in H$, $u' = \overline{-u}$ with $\overline{u}$ as in proposition \ref{pr:nuCTG} and $U' = \{ u' \talque u \in U \}$, $\# U = \# U'$.
Let $U \subset H$ and $U^\perp = H \setminus U' = (H\setminus U)'$, then the dual code of $\code_U$ is $\code_U^\perp = \code_{U^\perp}$
\end{thm}

\pf

Let $u, v \in H$, then one has that $\langle (\alpha^{\langle u,i \rangle})_{i \in H} , (\alpha^{\langle v,i \rangle})_{i \in H} \rangle =
\sum_{i \in H} \alpha^{\langle u+v,i \rangle }$

$$\sum_{i \in H} \alpha^{\langle u+v,i \rangle }= \sum_{i \in H} \alpha^{\langle \overline{u+v},i \rangle }=\left\{ \begin{array}{ll}
\frac{q (q-1)}{2}(\sup(\overline{u+v})   )  = 0  & ~ \mathrm{if} ~ \overline{u + v} \neq 0\\
\nonumber & \nonumber\\
\sum_{i \in H} 1 = (-1)^r   & ~ \mathrm{if} ~ \overline{u + v} = 0
\end{array} \right.$$where $\sup(\overline{u+v})$ is the number of nonzero coordinates of $\overline{u + v}$.

Then $ \langle \ev(Y^u) , \ev(Y^v) \rangle = 0$ for $u \in U$, $v \in U^\bot$ since $\overline{u+v} \neq 0$. On account of the dimension of
$\ff[U]$ and $\ff[U^\bot]$ and the linearity of the codes the proof is completed. \qed

The previous result shows that the dual of a toric code $\code_{P_1}$ is a toric code only when there is a convex polytope $P_2$ such that
$\overline{P_1}^\bot = \overline{P_2}$. However the dual of a generalized toric code is a generalized toric code.

\begin{rem}
The main results of this paper were published without proofs in \cite{ru2}. Later, a similar result to theorem \ref{te:me} has been obtained
independently in \cite{bras}.
\end{rem}

\vspace{0.2cm}

Summarizing, the matrix $M$ of the evaluation map $\ev: \fq[H] \to \fqn$ is

$$
M=\left( \begin{array}{ccccc} \alpha^{\langle u_1 , i_1 \rangle} & \alpha^{\langle u_1 , i_2 \rangle} & \cdots
& \cdots & \alpha^{\langle u_1 , i_n \rangle}  \\
\alpha^{\langle u_2 , i_1 \rangle} & \alpha^{\langle u_2 , i_2 \rangle} & \cdots
& \cdots & \alpha^{\langle u_2 , i_n \rangle} \\
\vdots & \vdots & \vdots & \vdots & \vdots \\
\vdots & \vdots & \vdots & \vdots & \vdots \\
\alpha^{\langle u_n , i_1 \rangle} & \alpha^{\langle u_n , i_2 \rangle} & \cdots & \cdots & \alpha^{\langle u_n , i_n \rangle}
\end{array}\right)
$$ where $\{ u_1, \ldots, u_n \} = \{i_1, \ldots, i_n\} = H$ and if moreover $u_j =
i_j$ then $M$ is a symmetric matrix, therefore we assume $u_j = i_j ~ \forall j =1, \ldots n$.

We have thus proved that a generator matrix of the code $\code_U$ with $U \subset H$, $k = \# U$, is the $(k\times n)$-matrix $M(U)$ consisting
in the $k$ rows $\alpha^{\langle u , i_1 \rangle},  \ldots, \alpha^{\langle u , i_n \rangle}$ of $M$ with $u \in U$ and a control matrix of
$\code_U$  is the $(n-k \times n)$-matrix $M(U^\bot)$ consisting of the $n-k$ rows $\alpha^{\langle u , i_1 \rangle}, \ldots, \alpha^{\langle u
, i_n \rangle}$ of $M$ with $u \in U^\bot$. Or equivalently the transpose of a control matrix is the $(n \times n-k)$-matrix consisting of the
$n-k$ columns $\alpha^{\langle u_1 , i \rangle}, \ldots, \alpha^{\langle u_n , i \rangle}$ of $M$ with $i \in U^\perp$ since we assume $u_j =
i_j ~ \forall j =1, \ldots n$.

The knowledge of the dual of a generalized toric code provides the following result to compute the minimum distance. This proposition is an
analogue of \cite[Proposition 2.1]{li} for toric codes whose proof remains valid for generalized toric codes. Using the control matrix one
simplifies the computations with respect to the generator matrix.

\begin{prop} \label{pr:medi}
Let $U \subset H$ and set $d$ an integer greater than or equal to 1. Suppose that $\forall ~ S \subset H$ with $\# S = d-1$ exists $V \subset
U^\perp$ with $\# V = d-1$ such that the square submatrix $M(S,V)$ of $M$ has nonzero determinant then $d( \code_U) \ge d$, where $M(S,V)$ is
the submatrix of $M$ corresponding to the rows of $S$ and columns of $V$, i.e. $M(S,V)=(\alpha^{\langle u_S , i_V \rangle})_{u_S \in S, i_V \in
V}$.
\end{prop}

\pf

The minimum distance of a linear code is greater than or equal to $d$ if any $d-1$ columns of a control matrix are linearly independent. A
control matrix of $\code_U$ is $M(U^\bot)$. Therefore the minimum distance of $\code_U$ is greater than or equal to $d$ if any $d-1$ columns of
$M(U^\bot)$ are linearly independent that is equivalent to the fact that exists a square submatrix of $M(U^\bot)$ with size $d-1$ and nonzero
determinant. \qed

Let  $\sigma (u) = u'$, and since $\sigma^2 = \mathrm{Id}$, one has that $\sigma$ is an involution. Moreover we order the elements of $H$ in
such a way that the matrix of the involution $\sigma$ has a characteristic form. By theorem \ref{te:me} we have that $B(\ev(Y^u),\ev(Y^v)) = 0$
if and only if $\overline{u + v} \neq 0$. We consider first the elements $u \in H$ such that $\sigma (u) = u' = u$, then $\overline{u + u} = 0$
and we have $B(\ev(Y^u),\ev(Y^u))=(-1)^r$ and $B(\ev(Y^u),\ev(Y^v))=0$ for all $v \in H \setminus \{u\}$. Then, we consider in $H$ the pairs of
elements $u$ y $\sigma(u)=u'$, with $u \neq \sigma(u)$, then $\overline{u + u'}=0$ and we have $B(\ev(Y^u),\ev(Y^{u'}))=(-1)^r$,
$B(\ev(Y^u),\ev(Y^v))=0$ for all $v \in H \setminus \{u'\}$ and $B(\ev(Y^{u'}),\ev(Y^v))=0$ for all $v \in H \setminus \{u\}$. Let $H=\{u_1 ,
\ldots , u_n \}$ ordered in the previous way. One has that the matrix $I_\sigma$ of the involution $\sigma$ is

$$
(-1)^r I_\sigma = \left( \begin{array}{llllllll}

1 & \nonumber  & \nonumber & \nonumber  & \nonumber & \nonumber & \nonumber  & \nonumber\\

\nonumber & \ddots & \nonumber & \nonumber & \nonumber  & \nonumber & \nonumber   & \nonumber\\

\nonumber & \nonumber  & 1 & \nonumber  & \nonumber & \nonumber & \nonumber  & \nonumber\\

\nonumber & \nonumber & \nonumber & 0 & 1 & \nonumber & \nonumber  & \nonumber\\

\nonumber & \nonumber & \nonumber & 1 & 0 & \nonumber & \nonumber  & \nonumber\\

\nonumber & \nonumber & \nonumber & \nonumber & \nonumber  & \ddots & \nonumber & \nonumber\\

\nonumber & \nonumber & \nonumber & \nonumber & \nonumber  & \nonumber   & 0 & 1\\

\nonumber & \nonumber & \nonumber & \nonumber & \nonumber  & \nonumber   & 1 & 0\\

\end{array} \right)
$$and therefore  $M^t M = (-1)^r I_\sigma$, and since $M^t = M$ one has that$$ M^{-1} = (-1)^r I_\sigma M$$

With these notations, the number of 1's in the main diagonal of the matrix $(-1)^r I_\sigma$ is established by our next proposition. Also, we
deduce that there are no self-dual generalized toric codes.

\begin{prop}
Let $\sigma$ be the involution $\sigma(u) = u'$ in $H$. The number of elements $u \in H$ such that $\sigma (u) = u$ is $2^r$ if $q$ is odd and
$1$ if $q$ is even. Moreover, there are no self-dual generalized toric codes.
\end{prop}

\pf

Let $u=(u_1, \ldots, u_r)$ in $H$, $\sigma (u) = u$ if and only if $2u_i = 0 \mod (q-1)$, for $i=1, \ldots, r$.

If $q$ is odd, then $2u_i = 0 \mod (q-1)$ if and only if $u_i$ is equal to $0$ or $(q-1)/2$. Therefore there are $2^r$ elements in $H$ with
$\sigma (u) = u$. We turn to the case $q$ even, then $q-1$ is odd and the only element in $H$ such that $2u_i = 0 \mod (q-1)$ for all $i$ is
$(0, \ldots, 0)$.

A linear code is self-dual if $\code^\bot = \code$, in particular $n$ must be even and $k=n/2$. If $q$ is even one has an odd length $n=(q-1)^r$
and therefore there are no self-dual toric codes with $q$ even. Let $q$ be odd, since there are $u_1 , \ldots u_{2^r} \in H$ such that $\langle
\ev(Y^u_i),\ev(Y^u_i)\rangle \neq 0$ the maximum dimension of a self-orthogonal code ($\code^\bot \subset \code$) is $n/2 - 2^{r-1} < n/2$, and
therefore there are no self dual generalized toric codes. \qed

\begin{exmp}
Let $\mathbb{F}_5$ the finite field with 5 elements and $r=2$. Therefore $H = \{0,1,2,3\} \times \{0,1,2,3 \}$. The length of a generalized
toric code $\code_U$ with $U \subset H$ is $n=4^2=16$.

We order the elements of $H$ to obtain $I_\sigma$ in the previous way. Since the base field has 5 elements one has $\sigma(u) = u$ for $2^2=4$
elements $u_1=(0,0)$, $u_2=(2,0)$, $u_3(0,2)$ and $u_4=(2,2)$. For the other elements of $H$ we have $\sigma (u) \neq u$ and we consider $u_j =
u$ and $u_{j+1} = \sigma(u)$, for instance $\sigma(0,1) = (0,3)$ and $\sigma(0,3)=(0,1)$. Therefore we write $u_5=(0,1)$, $u_6 = (0,3)$, $u_7 =
(1,0)$, $u_8 = (3,0)$, $u_9 = (1,1)$, $u_{10} = (3,3)$, $u_{11} = (1,2)$, $u_{12} = (3,2)$, $u_{13} = (1,3)$, $u_{14} = (3,1)$, $u_{15} =
(2,1)$, $u_{16} = (2,3)$. Let $i_j = u_j ~ \forall j \in \{1 , \ldots n \}$. This ordering of $H$ is not unique.

\vspace{0.3cm}

The evaluation matrix $M$ of the map $\mathbb{F}_5 [H] \to \mathbb{F}_5^n$ in the previous basis is

$$
M=\left(%
\begin{array}{cccccccccccccccc}
1 & 1 & 1 & 1 & 1 & 1 & 1 & 1 & 1 & 1 & 1 & 1 & 1 & 1 & 1 & 1 \\
1 & 1 & 1 & 1 & 4 & 4 & 1 & 1 & 4 & 4 & 1 & 1 & 4 & 4 & 4 & 4 \\
1 & 1 & 1 & 1 & 1 & 1 & 4 & 4 & 4 & 4 & 4 & 4 & 4 & 4 & 1 & 1 \\
1 & 1 & 1 & 1 & 4 & 4 & 4 & 4 & 1 & 1 & 4 & 4 & 1 & 1 & 4 & 4 \\
1 & 4 & 1 & 4 & 2 & 3 & 1 & 1 & 2 & 3 & 4 & 4 & 3 & 2 & 2 & 3 \\
1 & 4 & 1 & 4 & 3 & 2 & 1 & 1 & 3 & 2 & 4 & 4 & 2 & 3 & 3 & 2 \\
1 & 1 & 4 & 4 & 1 & 1 & 2 & 3 & 2 & 3 & 2 & 3 & 2 & 3 & 4 & 4 \\
1 & 1 & 4 & 4 & 1 & 1 & 3 & 2 & 3 & 2 & 3 & 2 & 3 & 2 & 4 & 4 \\
1 & 4 & 4 & 1 & 2 & 3 & 2 & 3 & 4 & 4 & 3 & 2 & 1 & 1 & 3 & 2 \\
1 & 4 & 4 & 1 & 3 & 2 & 3 & 2 & 4 & 4 & 2 & 3 & 1 & 1 & 2 & 3 \\
1 & 1 & 4 & 4 & 4 & 4 & 2 & 3 & 3 & 2 & 2 & 3 & 3 & 2 & 1 & 1 \\
1 & 1 & 4 & 4 & 4 & 4 & 3 & 2 & 2 & 3 & 3 & 2 & 2 & 3 & 1 & 1 \\
1 & 4 & 4 & 1 & 3 & 2 & 2 & 3 & 1 & 1 & 3 & 2 & 4 & 4 & 2 & 3 \\
1 & 4 & 4 & 1 & 2 & 3 & 3 & 2 & 1 & 1 & 2 & 3 & 4 & 4 & 3 & 2 \\
1 & 4 & 1 & 4 & 2 & 3 & 4 & 4 & 3 & 2 & 1 & 1 & 2 & 3 & 2 & 3 \\
1 & 4 & 1 & 4 & 3 & 2 & 4 & 4 & 2 & 3 & 1 & 1 & 3 & 2 & 3 & 2 \\
\end{array}%
\right)
$$

\vspace{0.3cm}

And we have that the matrix $M \cdot M^t = I_\sigma$ is

$$
I_\sigma=\left(%
\begin{array}{cccccccccccccccc}
1 & 0 & 0 & 0 & 0 & 0 & 0 & 0 & 0 & 0 & 0 & 0 & 0 & 0 & 0 & 0 \\
0 & 1 & 0 & 0 & 0 & 0 & 0 & 0 & 0 & 0 & 0 & 0 & 0 & 0 & 0 & 0 \\
0 & 0 & 1 & 0 & 0 & 0 & 0 & 0 & 0 & 0 & 0 & 0 & 0 & 0 & 0 & 0 \\
0 & 0 & 0 & 1 & 0 & 0 & 0 & 0 & 0 & 0 & 0 & 0 & 0 & 0 & 0 & 0 \\
0 & 0 & 0 & 0 & 0 & 1 & 0 & 0 & 0 & 0 & 0 & 0 & 0 & 0 & 0 & 0 \\
0 & 0 & 0 & 0 & 1 & 0 & 0 & 0 & 0 & 0 & 0 & 0 & 0 & 0 & 0 & 0 \\
0 & 0 & 0 & 0 & 0 & 0 & 0 & 1 & 0 & 0 & 0 & 0 & 0 & 0 & 0 & 0 \\
0 & 0 & 0 & 0 & 0 & 0 & 1 & 0 & 0 & 0 & 0 & 0 & 0 & 0 & 0 & 0 \\
0 & 0 & 0 & 0 & 0 & 0 & 0 & 0 & 0 & 1 & 0 & 0 & 0 & 0 & 0 & 0 \\
0 & 0 & 0 & 0 & 0 & 0 & 0 & 0 & 1 & 0 & 0 & 0 & 0 & 0 & 0 & 0 \\
0 & 0 & 0 & 0 & 0 & 0 & 0 & 0 & 0 & 0 & 0 & 1 & 0 & 0 & 0 & 0 \\
0 & 0 & 0 & 0 & 0 & 0 & 0 & 0 & 0 & 0 & 1 & 0 & 0 & 0 & 0 & 0 \\
0 & 0 & 0 & 0 & 0 & 0 & 0 & 0 & 0 & 0 & 0 & 0 & 0 & 1 & 0 & 0 \\
0 & 0 & 0 & 0 & 0 & 0 & 0 & 0 & 0 & 0 & 0 & 0 & 1 & 0 & 0 & 0 \\
0 & 0 & 0 & 0 & 0 & 0 & 0 & 0 & 0 & 0 & 0 & 0 & 0 & 0 & 0 & 1 \\
0 & 0 & 0 & 0 & 0 & 0 & 0 & 0 & 0 & 0 & 0 & 0 & 0 & 0 & 1 & 0 \\
\end{array}%
\right)
$$

\vspace{0.2cm}

Let $U=\{ (0,0), (1,0), (2,0), (0,1), (1,1), (2,1)$ and $\code_U$ the code defined by $U$ of length $n=16$ and dimension $k=6$. In this case
$\code_U$ is also a toric code  \cite[Theorem 2.5]{li} and \cite[example 5.1]{ru}. A generator matrix of $\code_U$ is the submatrix of $M$
consisting of the rows 1, 3, 5, 7, 9 and 15 of $M$. And a control matrix of $\code_U$, equivalently a generator matrix of $\code_U^\bot$, is the
submatrix of $M$ consisting of the rows 2, 4, 5, 7, 9, 11, 12, 13, 14 and 15 of $M$

\end{exmp}

\vspace{0.5cm}

\textbf{Acknowledgments:} The author thanks A. Campillo for helpful comments on this paper.

\bibliographystyle{hplain}
\bibliography{gtc-eacaAX}

\end{document}